# Magnetohydrodynamic simulation of the 2012-July-12 CME Event With the Fluxrope-G3DMHD Model

Chin-Chun Wu[1,a)]  Kan Liou[2,b)]  Brian Wood[1,c)]  Keiji Hayashi[3,d)]

[1] *Naval Research Laboratory, Washington, DC 20375, USA*
[2] *Applied Physics Laboratory, Johns Hopkins University, Laurel, Maryland, USA*
[3] *George Manson University, Fairfax, Virginia, USA*

[a)] Corresponding author: chin-chun.wu@nrl.navy.mil
[b)] kan.liou@jhuapl.edu
[c)] Brian.wood@nrl.navy.mil
[d)] khayashi@gmu.edu

**Abstract.** Coronal mass ejections (CMEs) and their driven shocks are a major source of large geomagnetic storms due to their large and long-lasting, southward component of magnetic field in the sheath and the flux rope (e.g., magnetic cloud). Predicting the strength and arrival time of southward fields accurately thus plays a key role in space weather predictions. To address this problem, we have developed a new model, which combines the global three-dimensional, time-dependent, magnetohydrodynamic (MHD), data-driven model (G3DMHD) and a self-contained magnetic flux-rope model [1]. As a demonstration and validation, here we simulate the evolution of a Sun-Earth-directed CME that erupted on 2012-July-12. The computational domain spans from 2.5 solar radii (Rs) from the surface of the Sun, where the flux rope is injected, to 245 Rs. We compare the time profiles of the simulated MHD parameters (Density, velocity, temperature, and magnetic field) with in situ solar wind observations acquired at ~1 AU by the *Wind* spacecraft and the result is encouraging. The model successfully reproduces the shock, sheath, and flux rope similar to those observed by *Wind*.

## INTRODUCTION

Geomagnetic storms play the major role in space weather. Due to their large and continuous southward fields, coronal mass ejections (CMEs) or, when in the solar wind, interplanetary coronal mass ejections (ICMEs) are the major source of large geomagnetic storms. In addition, geomagnetic storms tend to be severe when a CME drives a shock and forms a sheath between the shock and the magnetic ejecta (e.g., the magnetic cloud, MC) [e.g., 2,3]. This is because the sheath often contains large-amplitude, fluctuating southward magnetic fields, which can prime the magnetosphere or, when large enough, even induce a storm known as a two-step geomagnetic storm [*e.g.,* 4,5].

Most in-situ observations of the solar wind along the Sun-Earth-line is limited to the Sun-Earth system's L1 orbit, and restricts the prediction within the ~1 hour event transit time. This is often too short to have a practical application. A useful warning system needs a few days in advance. On the other hand, realistic global three-dimensional (3D), time-dependent magnetohydrodynamic (MHD) model may be a cheaper way to fulfill the requirement. Currently, only a few global 3D MHD simulation models are available to perform "realistic" solar wind simulation. The ENLIL model [6] is an operational solar wind model and has been adopted by multiple agencies, such as NOAA, Met Office, KSWC, and NASA's CCMC. However, validation of the performance of ENLIL has generally yielded modest results for solar wind parameter predictions [*e.g.*, 7]. Furthermore, ENLIL is a pure solar wind model. While it has been incorporated with pressure pulses (e.g., the Cone model) to simulate CME arrival time [e.g., 8], it cannot simulate ICME events with a magnetic cloud.

Therefore, it motivates us to develop a fluxrope CME model that builds directly into NRL (Naval Research Laboratory) in-house global 3D, time-dependent MHD simulation model (G3DMHD) [9,10]. Similar to ENLIL, G3DMHD is a pure solar wind model and has been used to stimulate non-cloud CMEs by inserting single pressure pulse to simulate a CME event or multiple pressure pulses to simulate multiple CME events [11]. The fluxrope model (EFR) is based on an analytic model of Chen [1]. The details about the fluxrope model is in preparation and will be submitted to and published elsewhere. Here we will demonstrate and validate the model by simulating a CME event that occurred on July 12, 2012.

## CME OBSERVATIONS

Figure 1 shows snapshot images of the CME eruption that occurred on July 12, 2012. *Cor1a* (of

SECCHI/STEREO) starts seeing a compression wave forming ahead at 16:25 UT on 2012-07-12 (see Figure 1a), and a CME loop at 16:55UT (Fig. 1b). CME images recorded by Cor2a, Cor2b, HI1b, and HI2b are showed in Fig.1c-f. Estimated propagation speed of the CME12 were (g) 690.8 (*Cor1a*), (h) 858.2 (*Cor1b*), (i) 1102 (*Cor2a*), (j) 1276 (*Cor2b*), (k) 1111.8 (*Hi1b*), and (l) 638.4 (*Hi2b*) km s$^{-1}$, respectively. The corresponding solar flare associated with this halo CME is a X1.4 flare peaked at 16:49UT from an active region (AR) 11520 at the heliographic coordinates S15W01 (https://cdaw.gsfc.nasa.gov/CME_list/halo). This event has been studied by several authors using observational or theoretical methods [11, 12, 13, 14, 15]. But NOAA only reports an X1.4 X-flare from AR11522 at N13W15 (https://www.ngdc.noaa.gov/stp/space-weather/solar-data/solar-features/solar-flares/x-rays/ goes /xrs/goes-xrs-report_2012.txt)

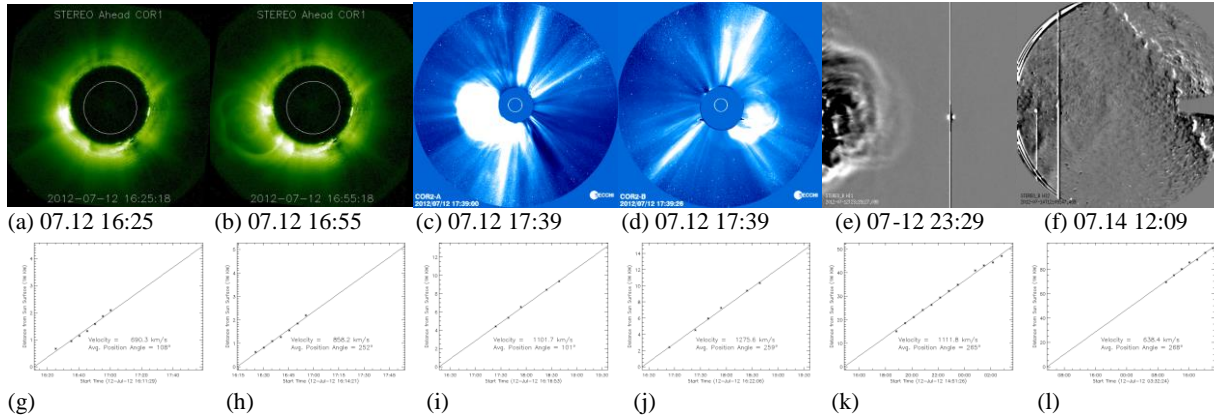

(a) 07.12 16:25  (b) 07.12 16:55  (c) 07.12 17:39  (d) 07.12 17:39  (e) 07-12 23:29  (f) 07.14 12:09

(g)  (h)  (i)  (j)  (k)  (l)

**FIGURE 1.** (a-f) CME images recorded by *STEREO/SECCHI* between 16:25UT on 2012-07-12 and 2012-07-14 with a source location at S15W01 for CME12 (https://cdaw.gsfc.nasa.gov/CME_list/halo/halo.html). (g-l) Estimated propagation speed of CME12 <V$_{CME12}$> are (g) 690.8 (*Cor1a*), (h) 858.2 (*Cor1b*), (i) 1102 (*Cor2a*), (j) 1276 (*Cor2b*), (k) 1111.8 (*Hi1b*), and (l) 638.4 (*Hi2b*) km s$^{-1}$ from FOV (field of view) of *Cor1a, Cor1b, Cor2a, Cor2b, HI1b, HI2b*, respectively.

## MODEL AND SIMULATION RESULTS

To simulate the July 12, 2012 CME event (CME12), we use a newly developed fluxrope-CME model that combines the G3DMHD solar wind model [9,10] and the EFR fluxrope model [1] (hereafter refer to EFR-G3DMHD). G3DMHD is a global three-dimensional, time-dependent, magnetohydrodynamic (MHD) numerical simulation model. The model solves a set of ideal-MHD equations (continuity, conservation of mass, momentum, and induction equations) using an extension scheme of the two-step Lax-Wendroff finite difference method [17]. The EFR model is a self-contained theoretical magnetic fluxrope model using an idealized partial toroidal flux rope with fixed footpoints [1]. In the model, the minor radius is assumed to be circular and does not deform. Expansion of a flux rope is based on self-similar growth of the initial flux rope up to the inner boundary of the G3DMHD computational domain (18 Rs for this study), where the G3DMHD model takes over [18]. Details about the EFR-G3DMHD model has been presented at EGU 2023 [18]. The MHD variables on the intersection between modeled rope moving at the constant translation and expansion velocities and the 18 Rs inner boundary surface are used as the time-dependent inner boundary values to perturb the background solar wind. Results of EFR-G3DMHD will be presented in this study.

Previous study suggests that CME12 was ejected from S11W15 [e.g., 12-16]. However, there was no report about the corresponding flare occurred in the southern hemisphere. On the other hand, an erupted flare was observed at N13W15 before the eruption of CME12. Therefore, we simulate three cases. Case 1 uses a pressure pulse to simulate the CME12 event with a source location at S11W15. Cases 2 & 3 use a flux-rope to simulate CME12 with the source location at N13W15 and S11W15.

Figure 2 shows snapshots of the solar wind radial speed ($v_r$) in the plane ($\theta = 7.5°N$) that contains the Earth's orbit from 18 to 345 $R_S$ at 2 different instances for the JUL12 event. Figure 2a & 2e (for time = 15:52UT) corresponds to the background solar wind speed ~0.62 hours prior to the CME12 eruption (16:25 UT). These velocity maps suggest that the heliosphere on the Earth plane is characterized by four magnetic sectors (indicated by the four blue spiral curves radiating outward from the Sun) during this rotation period. Here we determine the HCS (heliospheric current sheet) or SB (sector boundary) using $B_\varphi = 0$. About 41.45 hour later, the CME12-driven shock arrives at the Earth (See Fig. 2B, c, d, i, j, and k for the $v_r$ profiles and f, g, h, l, m, and n for the $Np$ profiles) at 09:52 UT on 2012-07-14 for all three cases at $\varphi = 0°$ (i.e., the Sun-Earth line).

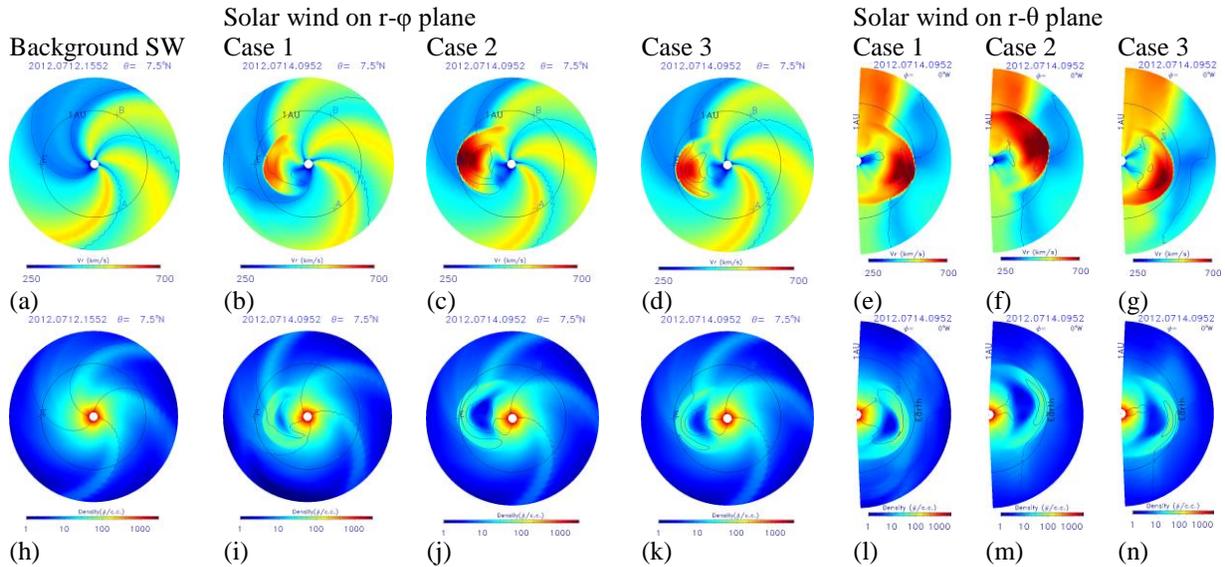

**FIGURE 2**. Temporal profile of simulated solar wind velocity on surface of angular cones that are centered at the Sun's center during 12-14 July by using velocity formula $Vr =150+500f_s^{-0.4}$ (km/s) for $Vr$ variation at 18 $R_\odot$. (a) & (h): Undisturbed solar wind velocity & density at conical angle at 7.5°N at 15:52UT on 2012-07-12. (b)-(g) & (j)-(n): evolution of $Vr$ & $Np$ for Case 1 (Fig. 2b,2c,2i,2l), Case 2 (Fig. 2c,2f,2j,2m), and Case 3 (Fig. 2d,2g,2k,2n) at 09:52 UT on July 14.

Comparison of the simulation results verses *in-situ* observations of *Wind* spacecraft for Case 1, 2, and 3 are presented in Figure 3a, 3b, and 3c (left, middle, and right panels), respectively. For Case 1: the simulation result has no clear shock, sheath, and driver (e.g., flux-rope). Note that while there is a jump in all parameters (A-D) near the observed shock, the magnetic field jump does not coincide with the other three parameter jumps. For Case 2: the temporal profiles of B, Tp, Np, and V are similar. The simulated sheath's duration is almost the same as the observed one and driver's size is surprisingly matched the observation. Furthermore, the V and Np are similar but Tp is clearly overestimate (same for the 3 cases). For Case 3, the temporal profiles of Bz and B show a better agreement than those of Cases 1 and 2 with the observations but the agreement in the B profile in the ejecta is not as good as in Case 1.

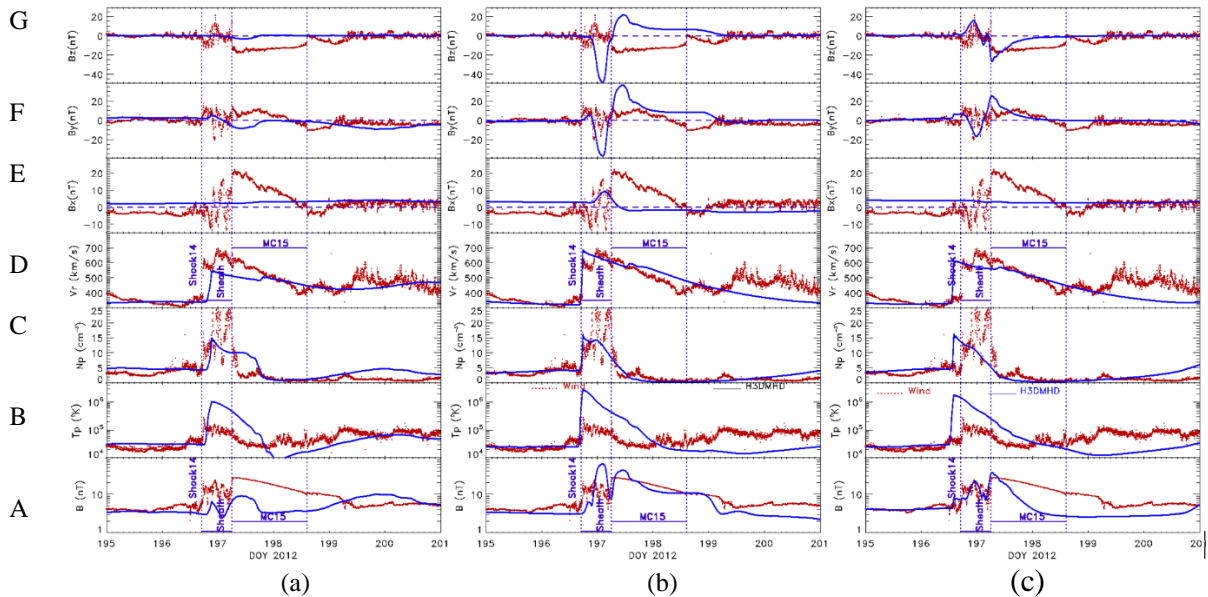

**FIGURE 3.** Comparison of simulation results (red-dots) verses *in-situ* observation (blue-curves). Comparison of in situ solar wind (red dots) profile and G3DMHD simulation results (blue traces) of CME12 for Case 1 (a, left panels), Case 2 (b, middle panels), and Case 3 (c, right panels). Panels from bottom to top show (A) total magnetic field, (B) proton temperature, (C) proton density, (D) the solar wind radial speed, (E) Bx, (F) By, and (G) Bz.

# DISCUSSION AND CONCLUTION

Our simulation results show that the profile of CME is sensitive to the perturbation type. A pressure pulse CME could simulate CME-driven shock's arrival time at the Earth correctly (see Figure 3A) but has no clear shock and sheath structures. When the CME is initiated with a flux-rope, the profiles of CME, including the shock, sheath, and ejecta, can be reasonably reconstructed (See Figure 3B & 3C). It is well-known that the CME source location is an important factor that affect the orientation of interplanetary magnetic fields (IMF) behind the shock (See Figures 3B/middle panels, 3C/right panels). The source locations of Cases 2 & 3 are different, N13W15 versus S11W15, respectively. For Case 2, the simulated IMF Bz switches sign in the sheath region and the initial turning in the driver matched well with the observations, but the duration of the driver is shorter than the observations. For Case 3, the simulated duration of the driver matches well with the observations, but the turning of simulated IMF Bz does not match the observations.

We have demonstrated the improvements of the G3DMHD in reconstructing solar wind and CME profiles at 1 AU by simulating the halo CME event on July 12, 2012. This is not to say the current state of the EFR-G3DMHD model is perfect. Some further studies are definitely required to understand the capability of the new model.

# ACKNOWLEDGMENTS

All data used in this study are obtained from the public domain. We thank the *Wind*, *STEREO* PI teams and the National Space Science Data Center at NASA/Goddard Space Flight Center, for providing the solar wind plasma and magnetic-field data. This work was supported partially by the Chief of Naval Research (BW & CCW), NASA grant of 80HQTR20T0067. The authors thank Drs. Christopher Kung and Sam Cable from Engility/DoD High Performance Computing Modernization Office PETTT program for his technical assistance in parallelizing the G3DMHD code.